\documentclass[twocolumn]{emulateapj}
\usepackage{graphicx}
\usepackage{bm}
\shorttitle{Fate of close-in exoplanet moons}
\shortauthors{F. Namouni}

\begin{document}        
\title{The fate of moons of close-in giant exoplanets} 
\author{Fathi Namouni}
\affil{Universit\'e de Nice, CNRS, Observatoire de la C\^ote d'Azur, BP 4229, 06304 Nice, France}
\email{namouni@obs-nice.fr}
\begin{abstract}
We show that the fate of moons of a close-in giant planet is mainly determined by the migration history of the planet in the protoplanetary disk. As the planet migrates in the disk from beyond the snow line towards a multi-day period orbit, the formed and forming moons become unstable as the planet's sphere of influence shrinks. Disk-driven migration is faster than the moons' tidal orbital evolution. Moons are eventually ejected from around close-in exoplanets or forced into collision with them before tides from the star affect their orbits. If moons are detected around close-in exoplanets, they are unlikely to  have been formed in situ, instead they were captured from the protoplanetary disk on retrograde orbits around the planets.
 \end{abstract}

\section{Introduction}
With the increasing detection of close-in exoplanets by precision transit photometry (87 planets as of this writing, Schneider 2010), the question of the possible presence of exomoon signatures is receiving much attention (Sartoretti and Schneider 1999, Han and Han 2002, Barnes and Fortney 2004, Szab\'o et al. 2006, Simon et al. 2007, Johnson and Huggins 2006, Kipping 2009a, 2009b). The likelihood of the presence of moons around close-in planets based on  dynamical arguments has been examined by a number of authors: Barnes and O'Brien (2002) studied the tides raised by a satellite on its close-in parent planet and showed that massive satellites do not survive tidal evolution as opposed to smaller albeit undetectable moons ($10^{-3}M_\oplus$ for periods smaller than 10 days). Domingos et al. (2006) analyzed satellite stability  in the conservative three-body problem and determined the outermost stable orbital distance of prograde and retrograde satellites. Cassidy et al. (2009) studied the tidal heating and mass loss from satellites around close-in planets. By using a much larger planet tidal dissipation factor than that used by Barnes and O'Brien (2002), they argued that Earth-size planetary moons may survive the age of the solar system around giant planets in close-in orbits.   In these earlier works, the effect of planet migration in the protoplanetary disk was not taken into account. Planet migration is a necessary step in the history of  close-in giant planets as they form outside the snow line where it is possible for ices to condense and for the planets to capture large amounts of gas from the protoplanetary disk (Sasselov and Lecar 2000, Garaud and Lin 2007).  Only after they formed, do they spiral inwards to meet their current orbits.  In this Letter, we examine the effect of planet migration on satellite stability and show that the most likely evolution outcome is satellite ejection from around the planet.  In section 2, we discuss how migration affects satellite stability and determine an analytical criterion for satellite ejection. In section 3, we confront the analytical criterion to a direct simulation of satellite evolution for a migrating Jupiter-like planet that formed beyond the snow line.  In section 4, we discuss the significance of our results and their possible extension.

\section{Planet migration and moon ejection}
Planetary satellites form in the circumplanetary nebula over time scales from $10^3$ to $10^7$ years (Canup and Ward 2002, 2006, Mosqueira and Estrada 2003a,b, Sasaki et al. 2010). The larger end of the scale describes late satellite formation that is required to explain the ice content of Jupiter's satellites and the partially differentiated interior of Callisto. Satellites also need to survive inward migration (Type I) forced by the circumplanetary nebula's torques in order to avoid collision with the planet. Those surving satellites must therefore have started forming when or ($\sim 10^2$ to $10^3$ years) before the parent gas giant planet opened a gap in the gas disk and significantly reduced the gas inflow inside its sphere of influence. Fully grown gas giants migrate in the protoplanetary disk at the rate of the viscous orbital decay of the disk (Type II migration, Ward 1997). The migration time is therefore comparable to or less than the disk's lifetime, of order a few $10^6$ years (Mamajek 2009). The migration time for a Jupiter-mass planet is $10^5$ years (Papaloizou and Terquem 2006). The typical time of orbital evolution owing to the tides raised by the formed or forming moon on the planet is a few (to several) $10^9$ years (Murray and Dermott 2000, Barnes and O'Brien 2002, Cassidy et al. 2009).  No significant tidal evolution of the moons therefore occurs during the planet's migration towards the star. As the migration time is much larger than the moon's period around the planet (1 to $10^2$ days for regular satellites of the solar system giants), the moon's orbit is not directly affected by the planet's migration (i.e. planet migration is an adiabatic change with respect to the satellite's revolution around the planet). However, the size of planet's sphere of influence (Hill sphere) decreases with the planet's orbital distance. The size of the Hill sphere is given as:
\begin{equation}
R_H=\left(\frac{M_p}{3M_\star}\right)^\frac{1}{3}\, a_p, \label{hill}
\end{equation}
where $M_p$  and $M_\star$ are the planet's and the star's masses respectively and $a_p$ is the planet's orbital semi-major axis around the star. Satellite orbits are stable in a zone whose radius is a fraction of the Hill radius $R_H$ that we denote $R_{SZ}=f R_H$. The factor $f\simeq 0.48$ for prograde satellites and $f\simeq 0.93$ for retrograde satellites of a Jupiter mass planet (Domingos et al. 2006).  The larger regular satellites of the three giant planets in the solar system have semi-major axes of order a few  $10^{-3}$ to $10^{-2}R_H$ well inside the stability zone (Table I).  As the planet migrates, the Hill radius, $R_H$, decreases with the planet's semi-major axis, $a_p$. A satellite initially located well inside the stability zone, becomes unstable when:
\begin{equation}
\frac{a_s}{R_H(t)}\sim f,   
\end{equation}
where the satellite's semi-major axis remains constant because of adiabatic invariance. This criterion determines the planet's semi-major at which it looses the  satellite. For a given prograde satellite semi-major axis the critical planet semi-major, $a_{p,{\rm crit}}$, is given as:
 \begin{equation}
a_{p,{\rm crit}}\sim \frac{a_s}{f} \ \left(\frac{3M_\star}{M_p}\right)^{\frac{1}{3}} \sim 0.06\left[\frac{a_p(t=0)}{3{\rm \, AU}}\right]\left(\frac{\alpha_s}{10^{-2}} \right) {\rm AU} \label{acrit}
\end{equation}
where $\alpha_s=a_s/R_H(t=0)$ is the ratio of the satellite's semi-major axis to the initial Hill radius (Table I), and where $f= 0.48$ has been used. 
If Jupiter had migrated inwards in the protoplanetary disk from its current position with its current satellite system, it would have lost the Galilean moons at $0.38$ AU (Callisto), at $0.22$ AU (Ganymede),  at $0.13$ AU (Europa), and at $0.09$ AU (Io). Titan would have left Saturn at $0.34$ AU if the planet had migrated inwards from $9.53$ AU. These estimates are indicative and do not account for (i) the possible destabilizing interaction of an unstable moon with the remaining stable moons prior to its ejection, (ii) the tidal evolution that the Galilean moons have been subjected to over the age of the solar system, (iii) the infalling gas on the planet during migration and (iv) the role of circumplanetary disk torques on the evolution  of the forming moons. These aspects are discussed in section 3 (i and ii) and section 4 (iii and iv).

\begin{table}
 \begin{minipage}{80mm}
  \caption{Dynamical parameters of solar system satellites. Semi-major axes are scaled to the Hill distance (\ref{hill}). Masses are scaled to that of the parent planet.}
  \begin{tabular}{@{}lccr@{}}
  \hline
   Planet &  Satellite & Semi-major axis& Mass\\
 \hline
 Jupiter & Io & 0.008&  4.70$\times 10^{-5}$\\
& Europa &0.012&  2.52$\times 10^{-5}$\\
& Ganymede & 0.020& 7.81$\times 10^{-5}$\\
& Callisto & 0.035&5.67$\times 10^{-5}$\\
Saturn & Mimas & 0.003 & 6.77$\times 10^{-8}$\\ 
& Tethys & 0.004&  1.09$\times 10^{-6}$\\
& Titan & 0.018 &  2.36$\times 10^{-4}$\\
& Iapetus & 0.054 &2.79$\times 10^{-6}$\\
Uranus &  Miranda & 0.002& 7.58$\times 10^{-7}$ \\
& Ariel & 0.003& 1.55$\times 10^{-5}$ \\
&Titania & 0.006&  4.06$\times 10^{-5}$ \\
& Oberon & 0.008&  3.47$\times 10^{-5}$\\ \hline
\label{table1}
\end{tabular}
\end{minipage}
\end{table}

\section{Numerical simulation}
To verify the migration-ejection instability of the previous section, we simulate the effect of planet migration on a hypothetical Jupiter and its Galilean satellites by integrating the full equations of motion of the system (sun, planet and satellites). The initial satellite semi-major axes would have to correspond to those before tides modified  the orbits over the age of the solar system. The standard semi-major axis variation owing to tides is $\dot a_s = C_p M_s a_s^{-11/2}$ where $C_p$ is a constant that depends on the planet's physical parameters (Murray and Dermott 2000). This variation allows us for a given initial semi-major axis of the innermost moon to derive the initial position of the remaining moons. Assuming Io had formed outside Jupiter's Roche radius ($\simeq R_p$, the planet's radius) before it moved to its current orbit ($\simeq 6 R_p$) shows that the other three satellites have started out their orbital evolution with semi-major axes smaller than their current values by less  than $0.4\%$. Tidal migration being small for the outer three moons, we choose to reduce the satellites' actual semi-major axes by: $50\%$ (Io), $0.4\%$ (Europa), and $0.15\%$ (Ganymede).  Callisto's orbital radius remains unchanged.  All satellites have initially circular orbits. The planet is set on a circular orbit at 5.2 AU. To simulate migration, we subject the planet to an additional force of the form $-k{\bm v}$. This leads to a semi-major axis evolution of the form $a_p(t)=a_{p0}\exp(-2k t)$ where $a_{p0}$ is the initial planet semi-major axis. The constant $k$ is chosen so that the planet reaches $a_p=0.04$ AU  in $10^5$ years. 

\begin{figure}
\includegraphics[width=87mm]{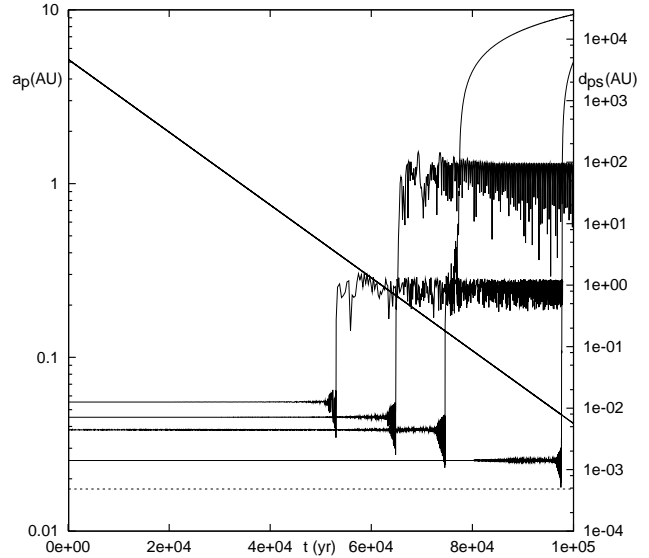}\\
\label{f11}
\caption{Planet migration and satellite ejection. The diagonal line shows the evolution of the planet's semi-major axis, $a_p$, (left scale, AU) from an initial value of 5.2 AU. 
The planet-satellite distances ($d_{ps}=|{\bm x}_p-{\bm x}_s|$ AU, right scale) of the four simulated satellites remain constant until the stability boundary closes in on them. The satellites are massless and have the initial semi-major axes (AU): $1.26\times 10^{-2}$ (Callisto), $7.11\times 10^{-3}$ (Ganymede), $4.44\times 10^{-3}$, (Europa) and $1.40\times 10^{-3}$ (Io). The lower dashed line at $4.81\times 10^{-4}$ AU (left scale) is the planet's radius.}
\end{figure}

We first check criterion (\ref{acrit}) by turning off mutual satellite interactions in the simulation. The results are shown in Figure (1) where the planet's orbital semi-major axis and the planet-satellite distance are shown as functions of time. The massless satellites are ejected from around the planet at $0.41$ AU (Callisto), at $0.23$ AU (Ganymede), at $0.14$ AU (Europa), and at $0.047$ AU (Io) in agreement with the expression (\ref{acrit}). Two satellites remain on star-bound orbits: Callisto with $a_s=0.8\,\mbox{AU}, \ e_s=0.45$, and Ganymede with $a_s=50\,\mbox{AU}, \ e_s=0.99$ where $e_s$ is the satellite's orbital eccentricity. Europa and Io, however,  are ejected from the planetary system. Whereas satellites are invariably ejected from around the planet, we find that their final orbits in the planetary system are sensitive to the initial orbital parameters as motion is chaotic  near the stability boundary. In agreement with the adiabatic character of planet migration with respect to the satellite's orbital motion, the satellite is oblivious to the planet's shrinking orbit until the stability boundary closes in on the satellite a few $10^3$ years before ejection. 

Mutual satellite interactions further disturb the system once instability closes in on the outermost satellite. Figure (2) shows how Ganymede's instability perturbs Europa and ejects it in its place. Mutual interactions result in an   earlier perturbation (a few $10^4$ years) before ejection than when mutual interactions were absent. This perturbation forces Io's orbit to assume a larger eccentricity that results in a collision with the planet. The remaining three satellites assume star-bound orbits: ($a_s=4.16\,\mbox{AU}, \ e_s=0.88)$ for Callisto, ($a_s=49.5\,\mbox{AU}, \ e_s=0.99)$ for Ganymede and ($a_s=3.12\,\mbox{AU}, \ e_s=0.91)$ for Europa. As with the massless satellite simulation, the instability outcome after ejection is sensitive to the initial conditions.

\begin{figure}
\includegraphics[width=87mm]{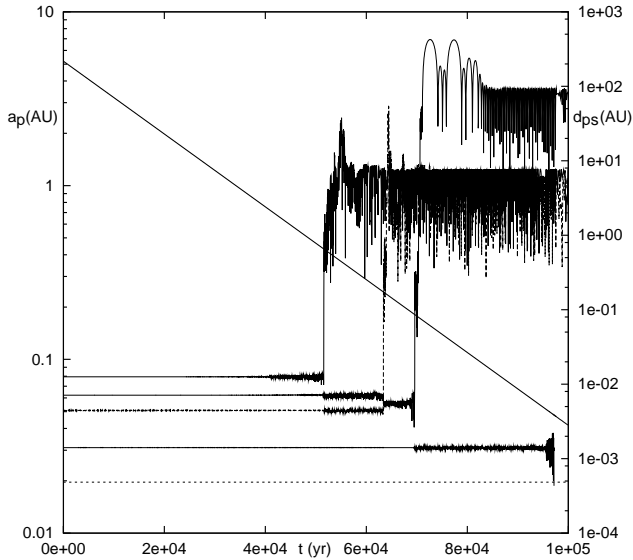}\\
\label{f11}
\caption{Planet migration and satellite ejection. The simulation is identical to that in Figure (1) except that mutual satellite interactions are turned on and satellites have the masses indicated in Table (1). Ganymede and Europa exchange orbits at the point where Ganymede must have been ejected forcing out Europa instead. Io is forced into a collision with the star because of its earlier perturbation by Ganymede.}
\end{figure}

\section{Conclusion}
This work illustrates how the migration-ejection instability operates for exoplanet satellites. In doing so we assumed that satellites form at semi-major axes in the range from $10^{-3}$ to $10^{-2}R_H$ as observed in the solar system. This assumption is reasonable as the three solar system planets with regular satellites (Table I) formed in different conditions at different orbital radii from the sun. In addition, their large regular satellites are likely to have been subjected to inward (Type I) migration by the circumplanetary disk's torques that resulted partly in the currently observed orbital resonances. A number of factors may alter the migration-ejection instability characteristics such as the ejection timescale and semi-major axis but they do not suppress it. For instance it is known that as the planet opens a gap in the disk, its growth may be significantly reduced but not completely halted (Kley et al 2001, Lubow et al 2002).  The accretion rate of order $10^{-2}$ to $10^{-4}M_\oplus$ yr$^{-1}$, depending on the circumstellar disk's physical properties, increases the size of the Hill sphere and may counter the effect of migration. As the Hill parameter depends on $M_p^{1/3}$, doubling the planet's mass during migration reduces the critical planet semi-major axis by 20\% only. The migration-ejection instability may also be mitigated if the planetary system's snow line is closer to the star than the nominal value of 3 AU  (Garaud and Lin 2007) allowing giant planet formation to occur closer to the star. Type I migration in the circumplanetary disk transports forming the satellites towards the planet and away from the stability boundary. In order for the satellites not to collide with the planet, the timing of the disk's dispersal needs to be linked to the satellites' arrival outside the Roche radius. Lastly,  eccentricity damping by the circumplanetary disk of forming satellite orbits (Ward 1988) would oppose the exciting effect of mutual  satellite interactions and make the migration-ejection instability effect similar to that on massless satellites.  As planetary satellites form outside the planet's Roche radius where differential tidal forces are not able to prevent planetesimal accumulation, a conservative estimate of the limiting planet semi-major axis, $a_{p,{\rm lim}}$, at which all moons are ejected is reached when the Roche radius,
 $R_R=({3M_p}/{2\rho_s})^{1/3}$ (for a rigid satellite), becomes comparable to the stability zone radius, 
 $R_{SZ}= fR_H$, where $\rho_s$ is the satellite's 
 mean mass density. It is given as:
 \begin{equation}
a_{p,{\rm lim}}\sim \left( \frac{M_\star}{M_\odot}\frac{2\,{\rm g\,cm}^{-3}}{\rho_s}\right)^\frac{1}{3} \left\{\begin{array}{ll}
0.016 \,{\rm AU}& \mbox{prograde}\\
0.008\, {\rm AU} &   \mbox{retrograde}
\end{array}\right.
\end{equation}
Notwithstanding the factors that mitigate the migration-ejection instability, it appears unlikely that the innermost moon that formed on a prograde orbit around a close-in giant exoplanet with $a_p\leq 0.02$ AU could survive the migration-ejection instability. If moons are observed by around close-in planets by precision photometry missions such as {\it Kepler} or {\it CoRot}, they are likely to be have been captured planetary embryos on retrograde orbits much like Triton around Neptune.

\acknowledgments{The author thanks the referee for useful comments on the manuscript.}

\end{document}